\documentclass[11pt]{article}

\usepackage{ }

\def\half{\textstyle{\frac{1}{2}}}

\def\S{\Sigma'}

\def\tint{{\textstyle\int}}
\def\hg{{\hat g}}
\def\hp{{\hat\pi}}

\def\s{\hskip.08em}

\def\b{\begin{eqnarray*}}  
\def\e{\end{eqnarray*}}    
\def\bn{\begin{eqnarray}}  
\def\en{\end{eqnarray}}   

\def\<{\langle}
\def\>{\rangle}

\def\no{\nonumber}

\def\{{\lbrace}
\def\}{\rbrace}
\def\pp{\partial}
\begin{document}  

\title{Quantum Gravity,  \\Constant Negative Curvatures,\\ and Black Holes}        
\author{John R. Klauder\footnote{klauder@phys.ufl.edu} \\
Department of Physics and Department of Mathematics \\
University of Florida,   
Gainesville, FL 32611-8440}
\date{ }
\bibliographystyle{unsrt}
\maketitle
\begin{abstract}
For purposes of quantization, classical gravity is normally expressed by canonical variables, namely the metric $g_{ab}(x)$ and the momentum
$\pi^{cd}(x)$. Canonical quantization requires a proper promotion of these classical variables to quantum operators, which, according to Dirac, the favored operators should be those arising from classical variables that formed
Cartesian coordinates; sadly, in this case, that is not possible. However, an affine quantization features promoting 
the metric $g_{ab}(x)$ and the momentric $\pi^c_d(x)\;[\equiv \pi^{ce}(x) \,g_{de}(x)]$ to operators. Instead of these classical variables belonging to a constant zero curvature space (i.e., instead of a flat space), they belong to a space of constant negative curvatures. This feature may even have its appearance in black holes, which could strongly point toward an affine quantization approach to quantize gravity.
\end{abstract}

\section{Basic Canonical and Affine Quantization}
\subsection{The essentials of canonical quantization}
A single classical momentum $p$ and coordinate $q$, with $-\infty <p,\,q<\infty$, and for which the Poisson bracket is
$\{q,p\}=1$, are promoted to self-adjoint operators, $p\rightarrow P$ and $q\rightarrow Q$,
which satisfy $[Q,P]=i\hbar$. The canonical coherent states are given by $|p,q\rangle\equiv 
\exp(-iqP/\hbar)\,\exp(ipQ/\hbar)\,|\omega\rangle$, where $(\omega Q+iP)|\omega\rangle=0$.
Stationary variations of normalized vectors $|\psi(t)\rangle$ of the quantum ($q$) action
 \bn A_q=\tint_0^T\langle\psi(t)|[i\hbar(\partial/\partial t)-\mathcal{H}(P,Q)]|\psi(t)\rangle\; dt \en
 lead to Schr\"{o}dinger's equation. The enhanced, with  $\hbar>0$ (as in the real world), classical ($c$) action  is given by
 \bn A_c\hskip-1.4em&&=\tint_0^T \langle p(t),q(t)|[i\hbar(\partial/\partial t)-\mathcal{H}(P,Q)]|p(t),q(t)\rangle\;dt \no \\
     &&=\tint_0^T[p(t)\dot{q}(t)-H(p(t),q(t))]\;dt. \en
     The connection of $H$ with $\mathcal{H}$ is given by 
    \bn H(p,q)\hskip-1.4em&&=\langle p,q|\mathcal{H}(P,Q)|p,q\rangle \no \\
     &&=\langle\omega|\mathcal{H}(P+p,Q+q)|\omega\> \no \\
     &&=\mathcal{H}(p,q)+\mathcal{O}(\hbar;p,q), \label{ll} \en
     which, when $\hbar\rightarrow0$, means $H(p,q)=\mathcal{H}(p,q)$. 
     Finally, we observe that the favored classical coordinates chosen to be
     promoted to operators, which, according to Dirac \cite{dirac} are Cartesian coordinates, is
     given by 
        \bn d\sigma_c(p,q)^2\equiv 2\hbar[|\!| \,d|p,q\>|\!|^2-|\<p,q|\,
     d|p,q\>|^2] =\omega^{-1}\,dp^2+\omega\,dq^2 \;.\label{ww} \en
     
     \subsection{The essentials of affine quantization}
     A single classical momentum $p$ and coordinate $q$, with $-\infty <p<\infty$ but now $0<q<\infty$, and for which the Poisson bracket is still  
$\{q,p\}=1$, are promoted to operators,  $p\rightarrow P$ and $q\rightarrow Q$, where $0<Q<\infty$ is self adjoint, but $P$ {\it can not} be self adjoint, i.e., $P^\dagger\neq P$. Instead of $p$, we choose to promote $pq\rightarrow (P^\dagger Q+QP)/2\equiv D$, a basic operator 
that is self adjoint, i.e., $D^\dagger=D$, and 
which satisfies $[Q,D]=i\hbar\,Q$. The affine coherent states, chosen for simplicity with $q$ and $Q$ as dimensionless, are given by $|p;q\rangle\equiv 
\exp(ipQ/\hbar)\,\exp(-i\ln(q)D/\hbar)\,|b\rangle$, where $[(Q-1)+iD/b\hbar]|b\rangle=0$ \cite{jk1}.

Stationary variations of normalized vectors $|\psi(t)\rangle$ of the quantum action
 \bn A'_q=\tint_0^T\langle\psi(t)|[i\hbar(\partial/\partial t)-\mathcal{H}'(D,Q)]|\psi(t)\rangle\; dt \en
 lead to Schr\"{o}dinger's equation. The enhanced, with $\hbar>0$, classical action is given by
 \bn A'_c\hskip-1.4em&&=\tint_0^T \langle p(t);q(t)|[i\hbar(\partial/\partial t)-\mathcal{H}'(D, Q)]|p(t);q(t)\rangle\;dt \no \\
     &&=\tint_0^T[-q(t)\dot{p}(t)-H'(p(t)q(t),q(t))]\;dt. \en
     The connection of $H'$ with $\mathcal{H}'$ is now given by 
 \bn H'(pq,q)\hskip-1.4em&&=\langle p;q|\mathcal{H}'(D,Q)|p;q\rangle \no \\
     &&=\langle b|\mathcal{H}'(D+pqQ,qQ)|b\> \no \\
     &&=\mathcal{H}'(pq,q)+\mathcal{O}'(\hbar;p,q), \label{yy} \en
     which, when $\hbar\rightarrow0$, means that $H'(pq,q)=\mathcal{H}'(pq,q)$. 
     Finally, we observe that 
     \bn d\sigma_a(p,q)^2\equiv 2\hbar[|\!| \,d|p;q\>|\!|^2-|\<p;q|\,
     d|p;q\>|^2]=(b\hbar)^{-1} q^2\,dp^2+(b\hbar)\,q^{-2}dq^2 ,\label{jj} \en 
     which does {\it not} lead to Cartesian coordinates.  Nevertheless.
     this metric is that of a {\it constant negative curvature}, whose value is $-2/ b\hbar$; see, e.g., \cite{tg,df}. That means that the affine metric in (\ref{jj}) is just as unique as the constant zero curvature (i.e., flat) metric of canonical quantization (\ref{ww})!\footnote{We note that the metric for spin variables coherent states leads to a constant {\it positive} curvature, the magnitude of which depends on the size of the  relevant Hilbert space.}
     
     \subsection{Comparison of canonical and affine quantization}
     The two versions of quantization described above apply to different problems. For example, assuming that $-\infty<p<\infty$, 
     the Hamiltonian of the harmonic oscillator, e.g., $H=(p^2+q^2)/2$, and $-\infty<q<\infty$, canonical quantization works and affine quantization fails, while when $0<q<\infty$, affine quantization works and canonical quantization fails \cite{jk1}.

The canonical story involves a constant zero curvature (i.e., a flat space), while the affine story involves a constant negative curvature, whose curvature value is $-2/b\hbar$. Such spaces are humanly visible only at one point, namely the `center point' where $q=1$ \cite{tg,df}. However, when a field is involved, it appears that a visible behavior can occur.

The favorable Cartesian coordinates of canonical quantization are different from the favorable affine coordinates of affine quantization. However, extremely close to $q=1$ the two versions are effectively the same. For a simple scalar field model, the metrics would be 
  \bn d\sigma_c^2=\tint[ A(x)^{-1}d\pi(x)^2+A(x)\,d\varphi(x)^2]\;dx \en
  for canonical, while that for affine is given by
\bn d\sigma_a^2=\tint [B(x)^{-1}\varphi(x)^2\,d\pi(x)^2+B(x)\,\varphi(x)^{-2}\,d\varphi(x)^2]\;dx.\en 
These curves have the center-point curvature (where $-2/B(x)$ denotes the value of the curvature).
It is noteworthy that each point $x$ has a visible center point where $\varphi(x)=1$, wherein its negative curvature at that $x$ is given by $-2/B(x)$. Thus, if $x$ is $1$ dimensional, this leads to visible spots, or perhaps a line. If $x=(x_1,x_2)$ is $2$ dimensional, this leads to a visible set of those points, or a region where $\varphi(x_1,x_2)=1$ with curvatures given by $-2/B(x_1,x_2)$, etc. 

This expression captures the central point of the constant negative curvature where it appears nearly flat, which in fact is the only point that can be realized in our sight of the negative curvature. This behavior holds for each point $x$ in space, thus enabling a field whose visible character is determined by the behavior of the non-dynamical term, $B(x)$.

Finally, observe that the canonical coherent states involve a single {\it additional} term of $q$ in the expression (\ref{ll}), while, for the 
affine coherent states, the expression for $q$ involves an {\it exponential} term in (\ref{yy}), passing from $\ln(q)$ to $q$, which entails a complete series. Thus, $q$ enters as a product term due to $D$ acting to `dilate' the result (hence the choice of symbol $D$). A similar exponential expression also emerges in the gravity coherent states; see \cite{gmcs}.

\section{An Affine Quantization of Gravity}
The basic story of affine quantization, and how it applies to gravity, has been presented in \cite{jk1} and several references therein; but if the reader wants additional foundations, we can recommend that particular article.
    

  Here we sketch an affine point of view for gravity. The classical momentric 
   field, $\pi^a_b(x)\,[\equiv \pi^{ac}(x)\,g_{bc}(x)]$, and the metric field, $g_{ab}(x)$, become the new basic variables, and these
   two variables have a joint set of Poisson brackets given by
   \bn &&\{\pi^a_b(x),\pi^c_d(x')\}=
   \half\,\delta^3(x,x')\s[\delta^a_d\s \pi^c_b(x)-\delta^c_b\s \pi^a_d(x)\s]\;,    \no \\
       &&\hskip-.20em\{g_{ab}(x), \s \pi^c_d(x')\}= \half\,\delta^3(x,x')\s [\delta^c_a g_{bd}(x)+\delta^c_b g_{ad}(x)\s] \;,      \\
       &&\hskip-.30em\{g_{ab}(x),\s g_{cd}(x')\}=0 \;. \no  \en
      
     Passing to operator commutations,  we are led by suitable coherent 
     states \cite{gmcs} to promote the Poisson brackets to the operators
 \bn   &&[\hp^a_b(x),\s \hp^c_d(x')]=i\s\half\,\hbar\,\delta^3(x,x')\s[\delta^a_d\s \hp^c_b(x)-\delta^c_b\s \hp^a_d(x)\s]\;,    \no \\
       &&\hskip-.10em[\hg_{ab}(x), \s \hp^c_d(x')]= i\s\half\,\hbar\,\delta^3(x,x')\s [\delta^c_a \hg_{bd}(x)+\delta^c_b \hg_{ad}(x)\s] \;, \\
       &&\hskip-.20em[\hg_{ab}(x),\s \hg_{cd}(x')] =0 \;. \no  \en
There are two irreducible representations of the metric tensor operator consistent with these
commutations: one where the matrix $\{\hg_{ab}(x)\}>0$, which we accept, and one where the matrix $\{\hg_{ab}(x)\}<0$, which we reject. 

   The classical Hamiltonian for our models is given \cite{adm} by
        \bn H(\pi, g)=\tint \{ g(x)^{-1/2} [\pi^a_b(x)\pi^b_a(x)-\half \pi^a_a(x)\pi^b_b(x)] 
           +g(x)^{1/2}\,^{(3)}\!R(x)\}\;d^3x, \en
 where $^{(3)}\!R(x)$ is the 3-dimensional Ricci scalar. For the quantum operators
we adopt a Schr\"odinger representation
 for the basic operators: specifically $ \hat{g}_{ab}(x)=g_{ab}(x)$ and 
   \bn \hat{\pi}^a_b(x)=-\half i \hbar\,[\,g_{bc}(x)\,(\delta/\delta\,g_{ac}(x))+(\delta/\delta\,
   g_{ac}(x)))\,g_{bc}(x)\,]\;.\en
 It follows that the Schr\"odinger equation is given by
                \bn && i\hbar\,\pp\,\Psi(\{g\}, t)/\pp t=\tint \{ [\hat{\pi}^a_b(x)\, g(x)^{-1/2} \,\hat{\pi}^b_a(x)-\half \hat{\pi}^a_a(x)\, g(x)^{-1/2} \,\hat{\pi}^b_b(x)] \no\\
         &&\hskip10em  +g(x)^{1/2}\,^{(3)}\!R(x)\}\;d^3x\;\Psi(\{g\}, t)\;, \label{rrr} \en
         where $\{g\}$ represents the $g_{ab}(x)$ matrix.
          
          \subsection{A closer look at the affine gravity metric} 
         Based  on the affine gravity coherent states \cite{gmcs}, the metric of
         favorable classical variables  is given by
          \bn &&\hskip-2em d\sigma_a(\pi,g)^2=\tint \{(b(x)\hbar)^{-1}\,g_{ab}(x)g_{cd}(x)\,
          d\pi^{bc}(x)\,d\pi^{da}(x) \label{zoom} \\ &&\hskip6em
          +(b(x)\hbar)\,g^{ab}(x)g^{cd}(x)\,dg_{bc}(x))\,dg_{da}(x)\}\;d^3\!x\;,  \no \en
          which involves constant negative curvatures at every point $x$. To make that point more clearly, we can arrange to rephrase (\ref{zoom}) as follows. 
               
           Let us introduce a strictly diagonal spatial metric given by
\bn g_{[]}(x)\equiv\left( \begin{array}{ccc}g_{[11]}(x)&0&0\\0&g_{[22]}(x)&0\\0&0&g_{[33]}(x) \end{array}\right ) \;, \en
with a determinant given by $\det\{g_{[]}(x)\}=g_{[11]}(x)\,g_{[22]}(x)\,g_{[33]}(x)$.
Next, we introduce three orthogonal matrices given by

 \bn O_1(x)\equiv \left( \begin{array}{ccc} 1&0&0\\0&C_1(x)&-S_1(x)\\0&S_1(x)&C_1(x) \end{array}
 \right) \en
 
 \bn O_2(x)\equiv \left (\begin{array}{ccc} C_2(x)&0&-S_2(x)\\0&1&0\\S_2(x)&0&C_2( x) \end{array}
\right) \en	

\bn 0_3(x) \equiv \left(\begin{array}{ccc} C_3(x)&-S_3(x)&0\\S_3(x)&C_3(x)&0\\0&0&1,\end{array}\right) \en where $C_a(x)\equiv\cos(\theta_a(x))$, $S_a(x)\equiv\sin(\theta_a(x))$, and $0\leq \theta_a(x)< 2\pi$.
The connection now is given by

\bn \{g_{ab}(x)\}=0_1(x)0_2(x)0_3(x)\,g_{[]}(x)\,0_3(x)^T0_2(x)^T0_1(x)^T \;, \en
where $T$ denotes transpose,
as well as its complete interchange given by

\bn g_{[]}(x)\equiv 0_3(x)^T0_2(x)^T0_1(x)^T\,\{g_{ab}(x)\}\,0_1(x)0_2(x)0_3(x) \;, \en
exploiting the fact that $0_a(x)0_a(x)^T=1=0_a(x)^T0_a(x)$ for each $a$.

This leads us to an alternative expression for (\ref{zoom}) given by
\bn \hskip-1.3em &&d\sigma_a(\pi,g)^2 \label{ppl} \\
   &&\equiv \tint\{(b(x)\hbar)^{-1}\,[g_{[aa]}(x)\,d\pi^{[aa]}(x)]^2+(b(x)
   \hbar)\,[g^{[aa]}(x)\,dg_{[aa]}(x)]^2\} \; d^3\!x. \no \en 
   We have chosen a common factor $b(x)$ and its inverse for all three terms. This is
   proper because the difference among each $g_{[aa]}(x) $ is simply their position along the matrix diagonal, which is a distinction with no physical significance.
   
    \section{Black Holes}
 Black holes are scattered around the universe and, loosely speaking, they tend to appear roughly similar. The
 appearance that pictures show is a cylindrical tube capped by a sprouting growth toward a
 flat topping; for sample pictures, see \cite{pic}. They can be visible due to stars being trapped
 in their presence. This section attempts to show that their configuration can be approximately
 seen via special points of their constant negative curvatures.
 
 \subsection{Special points on constant negative curvatures}
 Equation (\ref{ppl}) involves a wealth of potential constant negative curvatures as part of an affine
 gravity metric. They appear naturally different than Cartesian variables except at special
 points where $g_{[aa]}(x)=1=g^{[aa]}(x)$ for appropriate $x=(x_1,x_2,x_3)$. So far we have exploited 
 orthogonal matrices in order to present a more orderly expression for $d\sigma(\pi,g)_a^2$.
 However, we have not yet explored changes of the underlying space parameterized by $x$. To consider
 a change of coordinates, we first suggest a new set of special coordinates, $y=(y_1,y_2,y_3)$, that approximate a black hole in the equation 
 \bn y_1^2+y_2^2- e^{y_3} = 1\;. \label{yu} \en
 This equation illustrates a cylindrical tube of approximate radius $1$ when, e.g., $-6<y_3<0$, but this tube 
 blossoms out as $y_3$ passes through zero and rapidly requires an ever expanding cylindrical
 radius of $(y_1^2+y_2^2)^{1/2}$ as $y_3>0$.\footnote{The equation $y_1^2+y_2^2- e^{y_3} = C$, 
 where $1\leq C$ and $-6< y_3$, covers more of an idealization of a black hole and its surroundings.}
 
 We next develop a connection between the $x$ and the $y$ variables. We seek to have $\tilde{g}_{[aa]}(y)=1=\tilde{g}^{[aa]}(y)$, for all $[aa]$, when the three $y$ coordinates satisfy (\ref{yu}). This requirement suggests adopting the three diagonal metric terms, i.e., $g_{[aa]}(x)$ and $\tilde{g}_{[aa]}(y)$, and letting them act as `three-element vectors' such that $\tilde{g}_{[aa]}(y)=(\partial y_a/\partial x_b)\,{g}_{[bb]}(x)$, with a sum of $b$ with $[bb]$. Additionally, we change $\pi^{[aa]}$ to obtain
 $\tilde{\pi}^{[aa]}(y)=(\partial x_b/\partial y_a)\,\pi^{[bb]}(x)$, with a sum of $b$ with $bb$.
 Likewise, we introduce  $\tilde{g}^{[aa]}(y)=(\partial x_b/\partial y_a)\,g^{[bb]}(x)$. It follows that
 \bn \tilde{g}^{[aa]}(y)\,d\tilde{g}_{[aa]}(y)\hskip-1.5em&&=(\partial x_b/\partial y_c)\,(\partial y_c/\partial x_d)\,g^{[dd]}(x)\,dg_{[bb]}(x) \no \\
      &&=g^{[bb]}(x)\,dg_{[bb]}(x)\;, \label{wo1} \en
      properly summed. A similar relation is
      \bn \tilde{g}_{[aa]}(y)\,d\tilde{\pi}^{[aa]}(y)\hskip-1.5em
      && =(\partial y_b/\partial x_c)\,(\partial x_c
      /\partial y_d)\,g_{[dd]}(x)\,d\pi^{[bb]}(x) \label{wo2} \no \\
       &&=g_{[bb]}(x)\,d\pi^{[bb]}(x) \;, \en
       all of which leads to 
       \bn \hskip-1.3em &&d\sigma_a(\pi,g)^2 \label{pl} \\
   &&\equiv \tint\{(\tilde{b}(y)\hbar)^{-1}\,[\tilde{g}_{[aa]}(y)\,d\tilde{\pi}^{[aa]}(y)]^2+(\tilde{b}(y)
   \hbar)\,[\tilde{g}^{[aa]}(y)\,d\tilde{g}_{[aa]}(y)]^2\} \; d^3\!y\;. \no \en 
      
      We next exam the expression $\tilde{b}(y)\equiv b(x)$ for
      which the constant negative curvature is given by $-2/\tilde{b}(y)\hbar$.
      Assuming that one point of space, either denoted by $x$ or by $y$, 
      is the same as any other point of space, we are led to treat this
      term as similar, therefore, just as a constant, to be called $b$. Therefore,
      our final expression for the affine quantum metric is given by
      \bn \hskip-1.3em &&d\sigma_a(\pi,g)^2 \label{pll} \\
   &&\equiv \tint\{(b\,\hbar)^{-1}\,[\tilde{g}_{[aa]}(y)\,d\tilde{\pi}^{[aa]}(y)]^2+(b\,\hbar)
  \,[\tilde{g}^{[aa]}(y)\,d\tilde{g}_{[aa]}(y)]^2\} \; d^3\!y\;. \no \en 
  
  The choice of $y$ for the background space has been so as to secure that the values of
  $\tilde{g}_{[aa]}(y)=1=\tilde{g}^{[aa]}(y)$ when the three coordinate values obey the idealized 
  black hole behavior $y_1^2+y_2^2-e^{y_3}=1$, or a portion thereof. If that is so, then the results for (\ref{pll})
  have the appearance of being Cartesian coordinates, and thus may be seen as if they were actual
  Cartesian coordinates. Thus there would seem to be a sliver of nature that permits one to see
  starlight emitted along this `crack' in a black hole.

   \section*{Summary}
   The quantization of Einstein's version of classical gravity is possible using affine quantization
   instead of canonical quantization. Canonical quantization can not employ the promotion of classical
   Cartesian variables, as Dirac requires \cite{dirac}, because the classical gravity variables of phase space do not contain such variables \cite{john}. Fortunately, affine quantization lends itself toward fundamental affine variables that instead of a constant zero curvature (i.e., a flat space), have a constant negative curvature. An affine quantization of gravity has been partially developed 
   (see \cite{jk1,gmcs,bqg}) and it is ripe for additional analysis. In the present paper we have tried to bring the 
   affine gravity metric -- Eqs.~(\ref{zoom}), (\ref{ppl}), (\ref{pl}), and (\ref{pll}) -- into relation with physical expressions that
   are principally aimed at assigning such relations to black holes. While that aspect has been modestly treated, our analysis is open for further development.

\section*{Acknowledgements} Thanks to Ewa Czuchry for suggesting several clarifications of the
presentation.

   \end{document}